\documentclass[10pt,twocolumn]{IEEEtran}

\usepackage{filecontents}
\usepackage{amsbsy,amsmath,amssymb,epsfig,bbm,mathrsfs,fancyhdr,fancyvrb,url}
\usepackage{graphicx}
\usepackage{subfig}
\usepackage{graphicx}
\usepackage{float}
\usepackage{threeparttable}
\usepackage{breqn}

\usepackage{amsmath}
\usepackage{amsthm}

 \usepackage{multirow}
 \usepackage{graphicx}
 \usepackage[table,xcdraw]{xcolor}
 \usepackage[normalem]{ulem}

\usepackage{enumitem}
\usepackage{mathtools}
\usepackage[utf8]{inputenc}
\usepackage{multirow}
\usepackage{amsfonts}
\usepackage{epsfig}
\usepackage{amssymb}
\usepackage{graphicx}
\usepackage{amssymb,amsmath}
\usepackage{cite}
\usepackage{color,soul}
\usepackage{amsmath}
\usepackage{color}
\usepackage{lipsum}
\usepackage{dblfloatfix}
\usepackage{bbm}
\usepackage{multicol}
\usepackage{xcolor,colortbl}
\usepackage{algcompatible}
\usepackage{dblfloatfix}
\usepackage[noend]{algpseudocode}
\definecolor{Gray}{gray}{0.85}
\usepackage[subfigure]{tocloft}
\usepackage{epstopdf}

\newtheorem{re}{Remark}
\usepackage[colorlinks,bookmarksopen,bookmarksnumbered,citecolor=blue,urlcolor=blue]{hyperref}
\allowdisplaybreaks

\usepackage{amsmath, amsthm, amssymb}
\usepackage{bbding}
\usepackage[normalem]{ulem}
\usepackage{tabularx}
\usepackage{multirow}
\usepackage[normalem]{ulem}
\usepackage{array}

\usepackage{dblfloatfix}
\usepackage{multirow}
\usepackage{rotating}
\usepackage{tabularx}
\usepackage{array}
\usepackage{adjustbox,lipsum}
\usepackage{booktabs}
\usepackage{multirow}
\usepackage{tabularx}
\usepackage{array}
\usepackage{dblfloatfix}
\usepackage{blindtext}
\usepackage[normalem]{ulem} 
\usepackage{soul}
\usepackage[linesnumbered,ruled,vlined]{algorithm2e}
\SetKwInput{KwInput}{Input}
\SetKwInput{KwOutput}{Output}
\normalsize
\ifCLASSINFOpdf
\else
\fi
\usepackage{graphicx}
\usepackage{tikz}
\usepackage{bm}
\usepackage{supertabular}
\usepackage{longtable}
\usepackage{booktabs}

\begin{document}
	
		\title{E2E Migration Strategies Towards 5G: Long-term Migration Plan and Evolution Roadmap}	
	\author{\IEEEauthorblockN{Abulfazl Zakeri, \IEEEmembership{ Member, IEEE}, Narges Gholipoor, Mohsen Tajallifar, Sina Ebrahimi,  Mohammad Reza Javan, \IEEEmembership{Senior Member, IEEE}, Nader Mokari, \IEEEmembership{Senior Member, IEEE}, Ahmad Reza Sharafat, \IEEEmembership{Chair, ITU-D Study Group 2}
		}}			
	\maketitle
	\vspace{-2em}
	\begin{abstract}
After freezing the first phase of the fifth generation of wireless networks (5G) standardization, it finally goes live now and the roll out of the commercial launch (most in fixed 5G broadband services)
 and migration has been started.  
However, some challenges are arising in the deployment, integration of each technology, and the interoperability in the network of the communication service providers (CSPs). At the same time, the evolution of 5G is not clear and many questions arise such as whether 5G has
long-term evolution or when 5G will change to a next-generation one. This paper provides long-term migration options and paths towards 5G considering 
many key factors such as the cost, local/national data traffic, marketing, and the standardization trends 
in the radio access network (RAN), the transport network (TN),  the core network (CN), and E2E network.
Moreover, we outline some 
 5G evolution road maps emphasizing on the technologies, standards, and the service time lines. The proposed migration paths can be the answer to some CSPs' concerns about how to do long-term migration to 5G and beyond.
\end{abstract}
\begin{IEEEkeywords}
	Migration, 5G, evolution, roadmap, option, path.
\end{IEEEkeywords}  

\section{introduction}
The emergence of new communications services with diverse requirements necessitates the deployment of the fifth generation of wireless networks (5G) with an eye on its beyond for communication service providers (CSPs) such as telecom operators. ITU categorizes 5G services into three major classes, namely 1) enhanced mobile broadband service (eMBB), 2) ultra reliable and low latency communication (URLLC), and 3) massive machine-type communication (mMTC) that are specified in international mobile telecommunications-2020 (IMT-2020) documents as the overall requirements of 5G \cite{series2015imt}. 
 To meet these service requirements and businesses/networks flexibility/scalability, up to now, a plethora of academic research and standardization organizations have attempted to introduce and standardize several advanced technologies/protocols for the radio access network (RAN), the transport network (TN), and the core network (CN) for 5G. For example, the European telecommunications standards institute (ETSI) focuses on  network function virtualization (NFV), multi-access edge computing (MEC), and next generation protocols (NGP). In addition, 3rd generation partnership project (3GPP) deeply works on network slicing (NS) and defining the architectures and procedures for RAN and CN domains of a CSP network and their interworking procedures with earlier generations. However, there is a considerable gap between the introduction of these technologies and their deployment in practical scenarios. In addition to the technical considerations and requirements for the deployment and migration to the next generation, some other factors such as the investment cost, current status of the network/businesses, and the marketing trends should be considered.

The co-existence of some technologies in an end-to-end (E2E\footnote{In this article, by E2E, we mean the three parts of the telecom network and comprises RAN, TN, and CN.}) view may encounter some challenges such as interoperability and forward/backward compatibility. The deployment of these technologies in a network, which consists of various equipment belonging to different vendors, is very  challenging. An important step for migrating to 5G after completing its researching and standardization is finding an answer to how/why/where each CSP should deploy the technologies or integrate them in the existing network. This is the main issue that has not been completely addressed in the existing academic works on migration from 4G to 5G. In this regard, we provide several E2E migration options and paths from 4G towards 5G which is the main contribution of this paper.

\begin{figure*}[t]
	\centering
	\includegraphics[width=0.95\textwidth]{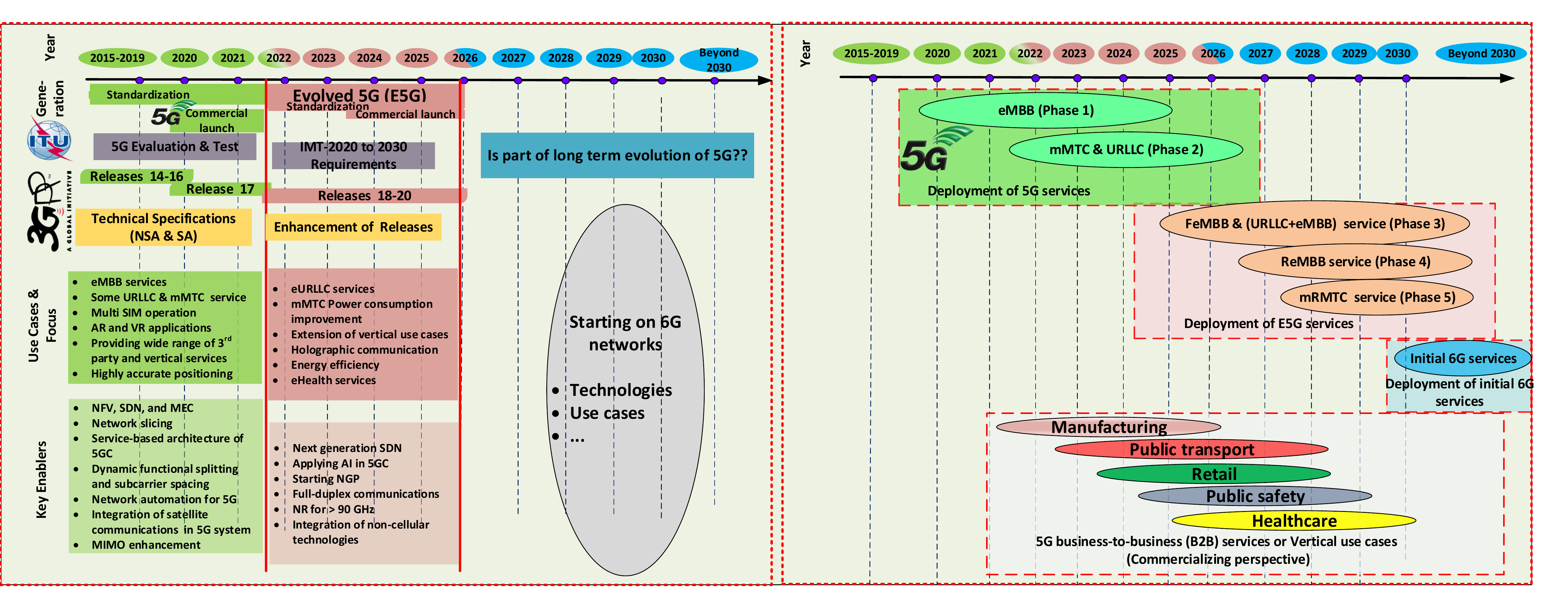}
	\caption{Right: 5G and E5G cellular networks roadmap, Left: commercializing 5G and E5G services roadmap. SDN: Softwared define network, NGP: next-generation protocols, AI: artificial intelligence, MEC: multi-access edge computing, eURLLC: enhanced URLLC,  FeMBB: further eMBB, ReMBB: reliable eMBB, RmMTC: reliable mMTC.
	}
	\label{Road_Map}
\end{figure*}

Although  the most major use case
of current 5G is eMBB, recently, some ever-new services such as telesurgery, telepresence, mixed reality, high secure/safe autonomous vehicles, and eHealth have been emerged in marketing \cite{li2018network}. These new services cannot be efficiently supported by the current 5G standards, and can be addressed in the evolution of 5G or 6G\cite{li2018network}. This requires that any operator have a plan for the evolution of 5G after/while deploying 5G,
especially from the perspectives of future-proof and network upgrade cost. Therefore, as another contribution of the paper, we provide an E2E vision to E5G. A holistic illustration of the proposed trend of 5G and its evolution considering the key enablers and use cases is shown in left-hand side of Fig. \ref{Road_Map}.
\textcolor{black}{
 As seen, 5G includes Releases 15-17 of 3GPP documents, E5G refers to Releases 18-20 around 2021-2025.
 The fundamentals of 5G are mostly deployed already and their enhancements are starting on 3GPP Releases 16 and 17 to completely meet IMT-2020 requirements. However, by considering new services that cannot be efficiently provided by the current 3GPP Releases, the beyond of 5G is needed. Hence, it is anticipated after around 2022, 5G evolution will be started to fully matured 5G in parallel with starting some fundamental studies on basic of 6G in standardization bodies. Note that this is similar to the evolution of 4G for compatibility and co-existing with 5G that has been started around 2014 in parallel with the starting standardization of 5G. Moreover, according to the Ericsson mobility reporting in  2019, it is expected mobile subscriptions by 4G will increase until 2021-2022.   
}

 The paper is organized as follows. We give the E2E-level migration towards 5G in Section \ref{E2E}. RAN migration stated in Section \ref{RAN Migration}
   gives the migration paths towards 5G-RAN, while the CN migration stated in Section \ref{CN Migration}  provides CN-level migration towards 5GC. Moreover, TN migration  stated in Section \ref{TN Migration} describes how to satisfy the TN-level requirements of 5G.
  Finally, the conclusion remarks are in Section \ref{conclusion}.

\textcolor{black}{
	\section{Long-term Migration Paths}
In this section, we discuss long-term viable migration paths for 5G from the E2E point of view. Regarding the deployment of 5G options standardized by 3GPP, in the early life of 5G, operators will have multiple options to migrate and deploy 5G as discussed in the following. 
}
\subsection{E2E Migration Paths} \label{E2E}
According to the 3GPP organization (specified in Releases 15 and 16), 5G deployment can be performed in five ways named options each with a specific architecture, different levels of investment cost and evolution cost, scalability, and support services key performance indicators (KPIs) and traffic. Hence, we provide a comprehensive comparison of options to express practical migration paths to 5G considering the current state of 4G. We mention that all the 5G options can be classified into two main categories: 1) standalone (SA) in which 5G RAN nodes, i.e., generation nodeB g(NB) handles joint data plane (DP)/user plane (UP) and control plane (CP)  and  2) non-standalone (NSA) in which gNB handles one of the UP or CP and the other one is handled by eNB \cite{GSMA}.

\subsubsection{{Fundamental comparison of 5G Options}}\label{5Goptions} 
The high-level descriptions mentioning the fundamental features of 5G options are as follows:
\\$\bullet~~$\textbf{Options 3/3A/3X:}
This 5G option has a series of features such as the new radio (NR) works in NSA with the anchor of eNB and CN is EPC. It supports eMBB services and 5G early devices. Moreover, it has a low investment cost to be deployed and needs RAN-level interworking between NR and eNB. Therefore, it  is already in use in some countries and many operators can select this for the first step towards 5G, especially for deploying 5G in mmWave band. Options 3X is the best for backhaul limited 4G networks, due to the possibility of splitting the total traffic in the gateway between eNB and NR. 
 \\$\bullet~~$\textbf{Options 7/7A:} 
  In this 5G option,  NR works in NSA with the anchor of eNB and CN is 5GC. This option supports eMBB services and some use cases of URLLC depending on the 5GC capabilities and as well as network slicing. Moreover,
   for deploy this option, operators needs to upgrade eNBs for supporting 5GC signaling (i.e., 5G non-access stratum (5G NAS))  and interfaces, and service-based 5GC with leveraging NFV and SDN. Therefore, it has more investment cost (in compared with option 3) and takes more time for the operators to deploy it. 
   From  the radio perspective, it is suggested for the high dense or hotspot areas  to deploy 5G NR in mmWave bands.
   \\$\bullet~~$\textbf{Options 4/4A:}  
    In this 5G option, NR works in NSA with the anchor of gNB and CN is 5G. This option supports flexible network slicing. It has a high investment cost to be deployed and more time to market (TTM). It needs to upgrade eNB for supporting 5GC signaling and interfaces. Moreover, operators in this 5G option can refarm some 4G bands and desire for the deployment of 5G in below $6$ GHz band. 
      \\$\bullet~~$\textbf{Options 2:}  
 This 5G option is the final target in which 5G can become standalone with E2E  5G network.
The main features of this option are that NR works in SA and CN is 5GC. It supports E2E network slicing. Hence, it has a high investment cost to be deployed and needs CN-level interworking with/without N26 interface \cite{thirdgeneration2017system}. In this option, operators can provide independent networks for vertical customers in an efficient and flexible manner.  It  is completely future-proof for E5G and evolution cost is low. 
        \\$\bullet~~$\textbf{Options 5:}  
  This 5G option has a series of features as eNB works in SA and CN is 5GC. It does not support mmWave (due to 4G's frequency bands) band and E2E network slicing. It can be deployed in some areas such as rural/urban for offering some low latency services  providing 5GC coverage. Moreover, it has low investment cost and needs to upgrade eNB for supporting 5GC NAS and its interfaces. It is not future-proof with E5G RAN. Therefore, operators can deploy it to provide flexible CN to support high-volume of 4G traffic.
 
 \textcolor{black}{
At the same time, it is possible that an operator deploys multiple options in different geographical locations for different scenarios and use cases (e.g., option 5 for some rural locations, and option 3 in milimeter-wave (mmWave) at  hotspot locations). Note that operators need to categorize the deployment locations as rural, urban, dense urban, and hotspot with the corresponding characteristics regarding to 3GPP standards \cite{22261}. 
}

\subsubsection{{Which one of the migration paths should be  selected?}}\label{Selecting_Migration_Paths}
	\textcolor{black}{Selection of a migration path depends on various key factors such as time to market (TTM), CAPEX/OPEX, future-proof of technologies, business trend, and the existing network conditions/architecture. We discuss them to provide the best path  for the operators with different marketing and technology status in Table \ref{Migration_Paths}.  Based on the technical and marketing comparison of different paths,  various benchmarking, and marketing analysis, we suggest path $1\rightarrow3\rightarrow4/2$ for tier-one operators because of marketing justifications and $1\rightarrow3\rightarrow7\rightarrow4/2$ for tier-2/3 operators to the nation-wide deployment of 5G.  This is because of in coverage of such operators, 5G supporting devices and 5G service demands are low at early stage and operators may have a concern about return of investment (ROI) in timing. Note that for some operators with the highest demands and subscriptions on 5G in some cities (e.g., high dense), $1\rightarrow4/2$ is suggested.
	}

As with many standardization organizations and vendors, the deployment of 5G will be spread on multiple phases \cite{GSMA}.
Regarding this, we consider three phases towards the maturity of 5G for operators: 1) early 5G (2018-2020), 2) full-scale 5G (2020-2023), 3) all-5G/E5G (beyond 2023-2026). It is anticipated that at the end of 2025, the study of standardization of  6G will be onset.  

\begin{table*}[]
	\centering
	\scriptsize
	\caption{Main migration paths towards nationwide deployment of 5G/E5G and these fundamental characteristics.}
	\label{Migration_Paths}
	\begin{tabular}{|l|l|l|}
		\hline
		\rowcolor[HTML]{32CB00} 
		\multicolumn{1}{|c|}{\textbf{Path} }                          & \multicolumn{1}{c|}{{{\textbf{Main factors and characteristics}}}       }                                                                                                                                                                                       & \multicolumn{1}{c|}{{\textbf{Considerations for operators}} }                                                                                                                     \\ \hline
		\cellcolor[HTML]{38FFF8}
		\textbf{1
			$\rightarrow$ 3$\rightarrow$7$\rightarrow$4/2} 
		& 
		\begin{tabular}[c]{@{}l@{}}
			$\bullet~$Operators can deploy 5G with the lowest CAPEX  investment\\ 
			$\bullet~$Operators could offer 5G eMBB services to customers\\ 
			$\bullet~$5G devices are available to support option 3\\ 
				\begin{tabular}[c]{@{}l@{}}
			$\bullet~$Due to the marketing and industries justification, highly\\ ~~~recommended from vendors and operators
			\end{tabular}
\\	$\bullet~$Utilized 4G as long time\\
		\end{tabular} 
		&
		\begin{tabular}[c]{@{}l@{}}
			$\bullet~$Operators should estimate their 5G early customers\\ 
			\begin{tabular}[c]{@{}l@{}}
				$\bullet~$The best for operators that have a 4G network widely and \\~~~aim to smoothly reach to SA 5G and develop towards E5G/6G
			\end{tabular}
			\\
			$\bullet~$Not appropriate for an operator who aim to provide 5G services\\ ~~~to vertical industrial in short TTM
			\\
			$\bullet~$Suitable for operators who aim to smoothly prepare 5GC, NFV, and SDN 
			\\$\bullet~$Suitable for tier 2/3 operators
			\\$\bullet~$Recommended for selecting nationwide 5G migration path
		\end{tabular} 
		\\ \hline
		\cellcolor[HTML]{38FFF8}
		\textbf{1$\rightarrow$3$\rightarrow$4/2}                                                     &                            
		\begin{tabular}[c]{@{}l@{}}
			$\bullet~$Operators can deploy early 5G with the minimum investment cost\\
			$\bullet~$Operators can refarm 4G spectrum after phase one of 5G\\ ~~~comparing \textit{Path} \textbf{1$\rightarrow$3$\rightarrow$7$\rightarrow$4/2} 
			\\ 
			\begin{tabular}[c]{@{}l@{}}$\bullet~$Operators target to start 5G with leveraging 4G \\~~~and directly expand 5G coverage \end{tabular} 
		\end{tabular}                      
		&  
		\begin{tabular}[c]{@{}l@{}}
			$\bullet~$ 
			Switching NSA to SA has a high cost\\
			$\bullet~$Recommended for operators with the rapid growth of 5G traffic\\ ~~~and customers (after around 2022) \\
			$\bullet~$Recommended for tier 1 operators\\
			$\bullet~$Operators need to expand 5G coverage rapidly
		\end{tabular} 
		\\ \hline
		\cellcolor[HTML]{38FFF8}
		\textbf{1$\rightarrow$4/2}       
		&   
		\begin{tabular}[c]{@{}l@{}}
			$\bullet~$Operators aim to deploy full-scale 5G \\
			$\bullet~$Have long TTM\\
			$\bullet~$Support all 5G services with enabling NS\\
			$\bullet~$NR works in  SA mode and CN is 5GC (all changes from 4G)
		\end{tabular}                                                                                                       &    
		\begin{tabular}[c]{@{}l@{}}
			$\bullet~$Operators should pay high investment cost\\
			$\bullet~$Operators need mature SDN, NFV, and E2E orchestrator in short time\\
			$\bullet~$May not have significant marketing revenue in start \\
			$\bullet~$Operators need to consider forward compatibility with Rel.16/17 \\
			$\bullet~$Recommended for special use cases (e.g., delay-sensetive applications) \\~~~and location (e.g., hotspot)
		\end{tabular}                                                                                                                                                                             \\ \hline
		\cellcolor[HTML]{38FFF8}
		\textbf{1$\rightarrow$7$\rightarrow$4/2}                                                     &                      
		\begin{tabular}[c]{@{}l@{}}
			$\bullet~$Support both 5GC/EPC NAS in starting 5G
			\\
			$\bullet~$Offer 5GC and NR capabilities
			\\ 
			$\bullet~$Utilize 4G RAN infrastructure and its spectrum in starting 5G 
			\\	$\bullet~$4G radio access is used as a anchor with deploying 5GC
		\end{tabular}                                                                                                        &          
		\begin{tabular}[c]{@{}l@{}}
			$\bullet~$Operators should pay more investment cost\\
			\begin{tabular}[c]{@{}l@{}}$\bullet~$Good path for operators whose have large deployed 4G with high traffic on 4G\\ ~~~coverage and high frequency band\end{tabular}\\
			$\bullet~$Good path for operators that have NFV, SDN, and 4G coverage 
			\\
			$\bullet~$Not Recommended for operators aiming nationwide 5G migration path
		\end{tabular}                                                                                                                            		
		\\ \hline
	\end{tabular}
\end{table*}


\section{ Radio Access Network Migration} \label{RAN Migration}
Regarding E2E migration paths that are in Section \ref{E2E}, the initial deployment of 5G RAN  is NSA and then goes to the SA mode. Migration to 5G RAN needs to be discussed from two main aspects: 1) 5G spectrum, 2) 5G RAN architecture, 5G base station (BS) implementation scenarios, and 5G RAN technologies. 
In the following, we explain and introduce the migration paths in each of these aspects.
\subsection{5G and E5G Spectrum Migration Paths}
What distinguishes 5G  from previous generations of cellular networks in the spectrum domain is the utilization of frequencies above $6$~GHz (mmWave frequency band). 
Despite the high path loss in this frequency range, a high bandwidth is available and the carrier bandwidth can reach $400$~MHz based on 3GPP reports \cite{3gpp.38.101-1}.

Generally, the 5G frequency spectrum is divided into three main categories: 1) the frequency spectrum below $1$~GHz, which is utilized for services that require deep coverage, e.g., IoT services, 2) frequency spectrum in the range of $1-6$~GHz \textcolor{black}{which} is suitable for services that require a high data rate and medium coverage, and 3) the mmWave frequency spectrum (above 6 GHz), which in spite of low coverage,  guarantees very high data rate for users. According to 3GPP reports, the first and the second categories are called FR1, and the third category is called FR2 \cite{3gpp.38.101-1}. 

\subsubsection{Important Considerations in Spectrum Migration  Paths}
In the following, we should consider the most significant factors in the spectrum migration plan:
\begin{itemize}
	\item \textbf{Operator marketing approach:} Each operator should prioritize the services based on some criteria such as community demands and the amount of revenue of services. Besides, it is necessary to consider the number of subscribers of each type of the requested services and their requirements.
	\item \textbf{Carrier aggregation (CA):} Each operator should provide the required spectrum according to the 3GPP CA standardization reports. 
	\item \textbf{5G end-user device trend:} 5G RAN implementation mainly depends on the availability of 5G devices that can support 5G RAN technologies such as CA and 5G frequency bands. In this regard, each operator should investigate the trends of 5G supporting devices and customers.
	\item \textbf{Regulatory approach:} Before deciding to provide equipment and spectrum, each operator should examine whether the spectrum bands are available or not in its region.
\end{itemize}

\subsubsection{Spectrum Migration Paths}
\begin{figure*}[t]
	\centering
	\includegraphics[width=1\textwidth]{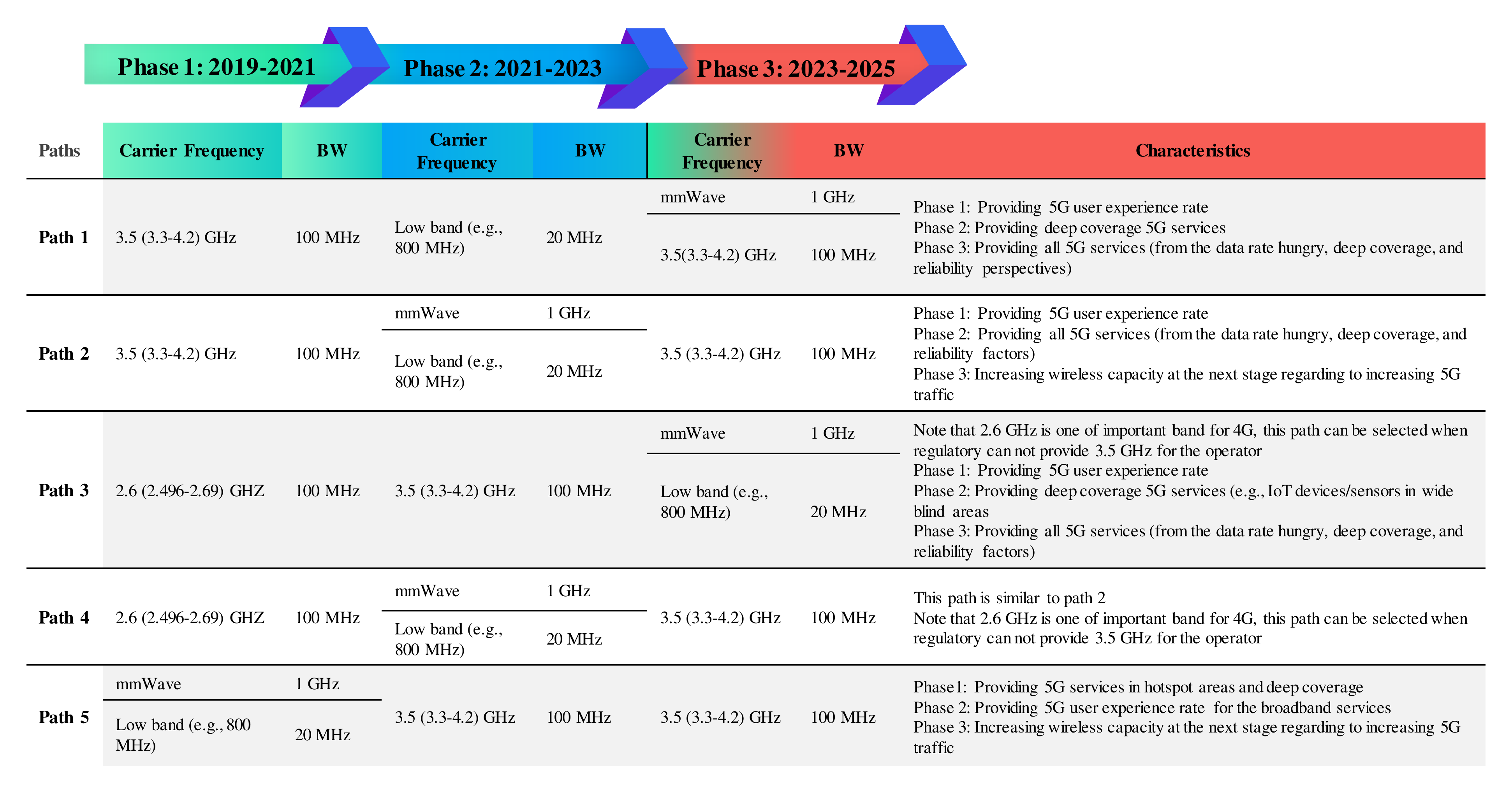}
	\caption{Long-term spectrum acquisition migration paths.}
	\label{SMP}
\end{figure*}

\textcolor{black}{Based on operators, standardization organizations such as 3GPP and global system mobile association (GSMA) reports, for implementing eMBB services in mid-band (3.5 GHz), about 80-100 MHz contiguous bandwidth should be used and in the high-frequency bands (mmWave) about $1$~GHz contiguous bandwidth is required. The reasons for the continuous bandwidth include: 1) reducing the user power consumption, 2) reducing the bandwidth waste, and 3) increasing the bandwidth efficiency. Since in some countries, enough bandwidth may not be available  in mid-band (3.5 GHz),   2.6 GHz band can also be used, because its bandwidth is about 100 MHz \cite{SpectrumGSMA,HuaweiGSMA}. It is worth noting that each operator can provide these spectra by re-farming legacy spectrum, sharing spectrum with other operators, or purchase new bands from regulatory. 
	Spectrum migration paths are shown in Fig. \ref{SMP}. Each operator can choose one of the proposed paths according to the mentioned considerations such as marketing plan and services.}

\subsection{5G RAN Architecture Migration Paths}
According to the 3GPP standardization reports, three different scenarios for NG-RAN deployment are proposed: 1) distributed-RAN (D-RAN): in which each gNB is deployed separately from other gNBs, 2) Co-located radio access technology (RAT): in which, both gNB and eNB are deployed on one site as part of a BS or each of which as a distinct BS, 3) centralized-RAN (C-RAN), in this scenario: the upper layer functions of RAN are implemented centrally \cite{3gpp.38.300}. 

In 5G, there are various options for centralizing functions and performing them in a centralized manner known as functional splitting (FS). The level of splitting and centralizing of these functions depends on the architecture and technologies of TN. As a result, gNB can be generally  separated into three units  as follows \cite{NGMNFS}:
\begin{enumerate}
	\item \textbf{Central unit (CU):} The CU performs the upper layer functions of gNB and controls one or more distributed units (DUs).
	\item \textbf{DU:}  The DU performs the lower layer functions of gNB, and each DU is controlled by one CU.
	\item \textbf{Radio unit (RU):} The RU contains the RF functionalities, i.e.,  PHY layer or the low level RAN functions.
\end{enumerate}

According to benchmarching (e.g., Heavy Reading  \cite{heavy2019transport}), operators' plan in the network architecture is a combination of C-RAN and D-RAN. This choice mainly depends on the availability of fiber (TN links), TN costs, and the services requested by users. Indeed, operators choose the C-RAN and D-RAN architectures based on the traffic in different regions, the type of services requested by users, and the status of their TN \cite{NGMNFS}. 
\subsubsection{Important Considerations in RAN Migration Paths}
\textcolor{black}{Deploying 5G RAN, depends on many aspects. The first aspect is determining
	the deployment in various geographical areas. In this regard, we divide geographical locations into four categories, called location scenarios (LSs), as follows:}
\begin{itemize}
	\item Targeted hotspot such as offices, stadiums, malls, and metro stations,
	\item Dense urban such as  city squares and commuter hubs,
	\item Urban such as residential areas and universities,
	\item Suburban and rural such as farms, inter-city roads, and villages.
\end{itemize}
In the FS scenarios, the locations of CU, DU, and RU are noticeable. These units can be co-located and distributed in the network regarding the mentioned RAN deployment scenarios. Therefore, the  deployment scenarios of these units are as follows \cite{NGMNFS}:
\begin{enumerate}
	\item RU, DU, and CU are located in  separate places. We show this scenario by RU+DU+CU.
	\item RU and DU are co-located and CU is located in a separate place. We show this scenario by (RU-DU)+CU.
	\item DU and CU are co-located and RU is located in separate place. We show this scenario by RU+(DU-CU).
	\item RU, DU, and CU are co-located. We show this scenario by (RU-DU-CU).
\end{enumerate}
Besides, some requirements in the network are needed to be considered to determine where the units should be located. These requirements are as follows \cite{itut2018tn5g}:
\begin{enumerate}
	\item \textbf{Requirements of RU-DU links (Fronthaul)}: The one-way delay between RU and DU should be less than $0.5$~ms. If the delay between the DU and RUs is more than $0.5$~ms, the radio access performance will be reduced. In addition to the delay requirement, these links should support high data rates at least $10$ gigabit (GB).
	\item \textbf{Requirements of DU-CU links (Midhaul)}: The one-way delay between CU and DU should be less than $3$~ms \cite{itut2018tn5g}.
\end{enumerate}
\begin{table*}[]
	\centering
	\begin{threeparttable}[b]
		\caption{Various unit placement scenarios (UPS).}
		\label{Table1}
		\begin{tabular}{|c |c|c|c|c|c|c|}
			\hline
			\rowcolor[HTML]{32CB00} 
			\textbf{\begin{tabular}[c]{@{}c@{}}Unit placement \\ scenarios (UPS)\end{tabular}} & \textbf{Access} & 
			\textbf{\begin{tabular}[c]{@{}c@{}}Pre-aggregation\\  Site\end{tabular}} & 
			\textbf{\begin{tabular}[c]{@{}c@{}}Aggregation \\ Site\end{tabular}} &
			\textbf{IP Core}
			& 
			\textbf{\begin{tabular}[c]{@{}c@{}}Fronthaul \\ challenges\end{tabular}} &
			\textbf{\begin{tabular}[c]{@{}c@{}}Midhaul \\ challenges\end{tabular}} \\ \hline
			\cellcolor[HTML]{38FFF8}
			\textbf{UPS 1} & RU, DU, CU & \_\_ & \_\_ & \_\_ & \_\_ & \_\_   \\ \hline\cellcolor[HTML]{38FFF8}
			\textbf{UPS 2} & RU, DU & CU & \_\_ & \_\_ & \_\_ & Low \\ \hline\cellcolor[HTML]{38FFF8}
			\textbf{UPS 3} & RU, DU & \_\_ & CU & \_\_ & \_\_ & Low \\ \hline\cellcolor[HTML]{38FFF8}
			\textbf{UPS 4} & RU, DU & \_\_ & \_\_ & CU & \_\_ & Moderate/ High\tnote{1}\\ \hline\cellcolor[HTML]{38FFF8}
			\textbf{UPS 5} & RU & DU, CU & \_\_ & \_\_ & Moderate & \_\_ \\ \hline\cellcolor[HTML]{38FFF8}
			\textbf{UPS 6} & RU & DU & CU & \_\_ & Moderate & Low \\ \hline\cellcolor[HTML]{38FFF8}
			\textbf{UPS 7} & RU & DU & \_\_ & CU & Moderate & Moderate/High \\ \hline\cellcolor[HTML]{38FFF8}
			\textbf{UPS 8} & RU & \_\_ & DU/CU\tnote{2} & \_\_ & High & \_\_\\ \hline\cellcolor[HTML]{38FFF8}
			\textbf{UPS 9} & RU & \_\_ & DU\tnote{2} & CU & High & Moderate/High \\ \hline
		\end{tabular}
		\begin{tablenotes}
			\item [1]{Depends on the distance between the aggregation site and the IP core sometime, it is  known as radio site gateway (RSG).}
			\item [2]{It is possible for geographical areas (cities) where access one-way delay to aggregation site is less than  $0.5$~ms, otherwise it is not possible.}
		\end{tablenotes}
	\end{threeparttable}
\end{table*}

\begin{table*}[]
	\caption{5G RAN architecture migration paths for different geographical locations with a fundamental comparison of them.}
	\label{tab:my-table}
	\resizebox{\textwidth}{!}{%
		\begin{tabular}{|c|c|l|l|l|l|}
			\hline
			\rowcolor[HTML]{34CDF9} 
			\textbf{Deployment Scenarios} & \textbf{Paths} & \multicolumn{1}{c|}{\cellcolor[HTML]{34CDF9}\textbf{Possible UPS}} & \multicolumn{1}{c|}{\cellcolor[HTML]{34CDF9}\textbf{Advantages}} & \multicolumn{1}{c|}{\cellcolor[HTML]{34CDF9}\textbf{Disadvantages}} & \multicolumn{1}{c|}{\cellcolor[HTML]{34CDF9}\textbf{Examples}} \\ \hline
			\rowcolor[HTML]{FFCC67} 
			\cellcolor[HTML]{FFCC67} & P1 & UPS 1, UPS 5 – UPS 7 & \begin{tabular}[c]{@{}l@{}}- Low Radio Equipment Cost\\ - Centralization Benefits\end{tabular} & - Moderate Fronthaul Challenge & \cellcolor[HTML]{FFCC67} \\ \cline{2-5}
			\rowcolor[HTML]{FFCC67} 
			\multirow{-2}{*}{\cellcolor[HTML]{FFCC67}\textbf{Indoor Hotspot}} & P2 & UPS 5- UPS 7 & - Low Radio Equipment Cost & - Fronthaul challenge (Lower than P1) & \multirow{-2}{*}{\cellcolor[HTML]{FFCC67}-Offices, stadiums, mall, metro station} \\ \hline
			\rowcolor[HTML]{00D2CB} 
			\cellcolor[HTML]{00D2CB} & P1 & UPS 1, UPS 5 – UPS 9 & \begin{tabular}[c]{@{}l@{}}- Low Radio Equipment Cost\\ - Centralization Benefits\end{tabular} & - High Fronthaul Challenge & \cellcolor[HTML]{00D2CB} \\ \cline{2-5}
			\rowcolor[HTML]{00D2CB} 
			\cellcolor[HTML]{00D2CB} & P2 & UPS 1 – UPS 4 & - Relaxed Fronthaul & \begin{tabular}[c]{@{}l@{}}- Higher Radio Equipment Cost than P1\\ - Less Centralization Benefits than P1\end{tabular} & \cellcolor[HTML]{00D2CB} \\ \cline{2-5}
			\rowcolor[HTML]{00D2CB} 
			\cellcolor[HTML]{00D2CB} & P3 & UPS 1, UPS 5 – UPS 9 & \begin{tabular}[c]{@{}l@{}}-  Less Fronthaul Challenge than P1\\ - Higher Centralization Benefits than P2\end{tabular} & \begin{tabular}[c]{@{}l@{}}-  Higher Radio Equipment Cost than P1\\ - Lower Centralization Benefits than P1\end{tabular} & \cellcolor[HTML]{00D2CB} \\ \cline{2-5}
			\rowcolor[HTML]{00D2CB} 
			\cellcolor[HTML]{00D2CB} & P4 & UPS 2 – UPS 4 & \begin{tabular}[c]{@{}l@{}}- Lower Radio Equipment Cost than P2\\ - Relaxed Fronthaul\end{tabular} & \begin{tabular}[c]{@{}l@{}}- Less Coverage than P1-P3\\ - Low Centralization Benefits\end{tabular} & \cellcolor[HTML]{00D2CB} \\ \cline{2-5}
			\rowcolor[HTML]{00D2CB} 
			\multirow{-5}{*}{\cellcolor[HTML]{00D2CB}\textbf{Dense Urban}} & P5 & UPS 5 – UPS 9 & \begin{tabular}[c]{@{}l@{}}-  Lower Radio Equipment Cost than P4\\ -  Higher Centralization Benefits than P4\end{tabular} & - Moderate\ High Fronthaul Challenge & \multirow{-5}{*}{\cellcolor[HTML]{00D2CB}- City squares, commuter hubs} \\ \hline
			\rowcolor[HTML]{EFEFEF} 
			\textbf{Urban} & P1 & UPS 1 – UPS 4 & - High Coverage & - High Radio Equipment Cost (Dependent on UPS) & - Residential areas, Universities \\ \hline
			\rowcolor[HTML]{FFFC9E} 
			\textbf{Rural} & P2 & UPS 1 & - High Coverage & - High Radio Equipment Cost (Dependent on UPS) & - Farm, Inter-city road \\ \hline
		\end{tabular}%
	}
\end{table*}

\begin{figure*}[t]
	\centering
	\includegraphics[width=1\textwidth]{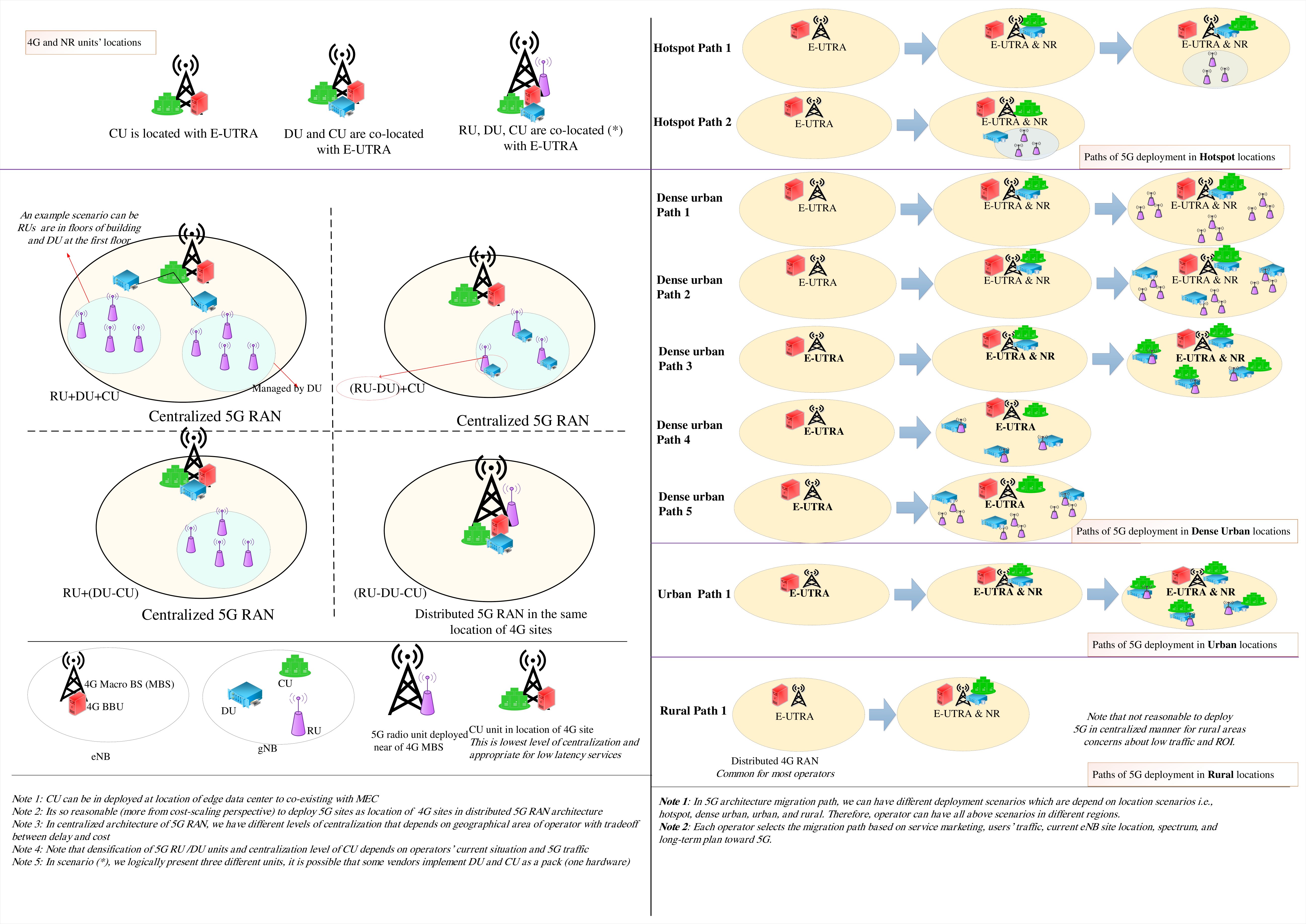}
	\caption{Left: all possible unit deployment scenarios of gNB, right: main migration paths for different deployment  areas.}
	\label{hotspot_path1}
\end{figure*}

Based on the benchmarking and business justifications, hotspot areas have priority for implementing 5G for operators. Next, 5G is implemented in dense urban areas, and after that in urban and rural areas which have lower population density.
It is worth noting that deployment paths depend on LSs. In the following, we analyze the migration path for D-RAN based networks which is deployed by many operators. It should be noted that in all of scenarios, NR is first co-located with E-UTRA (LTE RAN), and the operator first deploys NSA  (option 3). After implementing the NAS and preparing 5GC, the operator deploys SA (option 2). In the following, we analyze migration paths for four scenarios as follows:\\
\textbf{1-Hotspot/Indoor Hotspot}\\
In hotspot areas, small cells should be deployed to meet the high traffic requirements of those areas. In order to deploy small cells, we should consider the distance between these small cells and the macro cell, the feasibility of implementing fronthaul, and the requirements of fronthaul such as delay and throughput. If the fronthaul requirements are met between small cells and the macro cell, only RU unit is deployed in small cells, and the DU and CU can be deployed in the macro cell or farther. If it is not possible to implement fronthaul, we should deploy a DU near RUs to circumvent fronthaul's challenges.  In the following, we present two examples of scenarios of migration paths for the hotspot. It is worth noting that the path selection depends on the distance between hotspot area and macro site. At the same time, the operator should consider fronthaul challenges such as delay and throughput. \\
\textbf{Path 1:} This path is shown in Fig. \ref{hotspot_path1}. In this path, each operator deploys NR next to the E-UTRA in each macro cell. Then, they can deploy small cells that have RUs for the hotspot areas and DU and CU processing is done in the higher layers of the network such as macro cell. %
\\
\textbf{Path 2:} In this path, each operator deploys small cells that have RU and DU units for the hotspot areas and CU processing is done in the higher layers of the network such as macro cell. 
\\
\textbf{2- Dense Urban} \\In the urban areas, if traffic is high, the NR need be co-located with the 4G macro cells, otherwise it is not necessary to deploy NR in the macro cell in the first step. Similar to the hotspot scenario, to satisfy high traffic requirements of dense urban areas, small cells should be deployed in dense urban. Moreover, for selecting appropriate unit placement scenario (UPS)( as shown in Table \ref{Table1}), we should consider fronthaul challenges and service requirements in that area. In the following, we present two examples of scenarios of migration paths for the hotspot:\\
\textbf{Path 1:} This path for dense urban is similar to path 1 for the hotspot scenario as shown in Fig. \ref{hotspot_path1}. Thus, the operator deploys NR next to the E-UTRA in each macro cell. Then, they can deploy small cells that have RUs for dense areas and DU and CU processing is done in the higher layers of the network such as macro cell. 
\\
\textbf{Path 2:} This path for dense urban is similar to path 2 for the  hotspot scenario. Thus, the operator deploys NR next to the E-UTRA in each macro cell. Then, they can deploy small cells that have RU and DU units for dense areas and CU processing is done in the higher layers of the network such as macro cell.
\\
\textbf{3-Urban} \\For urban  areas, we only deploy small cells in areas where traffic congestion is high, depending on the type of service demands, and the fronthaul challenges and requirements;  DU, CU, and RU can be located in the macro cell, small cell, or between them  as shown in Fig. \ref{hotspot_path1}.
\\
\textbf{4-Rural} \\In the final phase of 5G implementation,  as shown in Fig. \ref{hotspot_path1}, NR equipment is added to  the 4G macro cell in rural areas and areas with low user density. 


\section{{Core Network Migration}}\label{CN Migration}
Throughout the past few years, enabling technologies such as NFV, SDN, MEC, and self-organizing networks (SONs) have been deployed more or less in the network of CSPs. These technologies are the key players in 5G, especially in the CN domain. We first review the high-level options in CN for migrating to 5G. Next, we introduce the possible states of CN architecture towards 5G. Finally, we provide two \textcolor{black}{examples of} migration paths for CN architecture.

\subsection{5G CN Options}
Regarding to standardization and benchmarking \cite{thirdgeneration2017system,GSMA}, three main options exists for the migration from 4G to 5G. Table \ref{highlevel_solutions} provides the fundamental comparison between these options. In this table, we \textcolor{black}{compare} three high-level CN options, namely EPC, EPC+, and 5GC with each other from virtualization, control/user plane separation (CUPS), and historical point of view.

\begin{table*}[]
	\caption{CN options with the corresponding key features \cite{thirdgeneration2017system,GSMA}.}
	\label{highlevel_solutions}
	\centering
	\scriptsize
	\begin{tabular}{|c|c|c|c|c|}
		\hline
		\rowcolor[HTML]{32CB00} 
		\textbf{Solution}                     & \textbf{Virtualization} & \textbf{Separation}                                           & \textbf{Year}                                                        & \textbf{Description}                             \\ \hline
		\cellcolor[HTML]{38FFF8}\textbf{EPC}  & Optional                 & Disabled                                                                         & \begin{tabular}[c]{@{}c@{}}Releases 8-13\\ (2008-2016)\end{tabular}  & \begin{tabular}[l]{@{}l@{}}- Reference Point Interfaces (Sx Interfaces)\\ - Using proprietary protocols (e.g., Diameter, eGTP,\\ \quad S1AP, and PFCP) for each interface\\ - {Possible deployment of DECOR 
		}\end{tabular} \\ \hline
		\cellcolor[HTML]{38FFF8}\textbf{EPC+} & Mandatory                & \begin{tabular}[c]{@{}c@{}}Featuring\\ 3GPP CUPS\end{tabular}                                                              & \begin{tabular}[c]{@{}c@{}}Releases 14-16\\ (2016-2019)\end{tabular} & \begin{tabular}[l]{@{}l@{}}- Reference Point Interfaces (Sx Interfaces \\ \quad including Sxa, Sxb, and Sxc due to CUPS)\\ - Using proprietary protocols (e.g., Diameter, eGTP-C,\\ \quad eGTP-U, S1AP, and PFCP) for each interface\\ - {Possible deployment of DECOR}\end{tabular}    \\ \hline
		\cellcolor[HTML]{38FFF8}\textbf{5GC}  & Mandatory                & \begin{tabular}[c]{@{}c@{}}Designed\\Separately\\from the\\beginning\end{tabular} & \begin{tabular}[c]{@{}c@{}}Releases 15-17\\ (2016-2021)\end{tabular} & \begin{tabular}[l]{@{}l@{}}- Service-Based Interfaces for CP function group\\ \quad (Nx Naming for interfaces (e.g. Nsmf, Namf))\\ \quad(Using HTTP/2-Based REST APIs)\\ - Reference Point Interfaces for UP NFs and their\\ \quad interconnections (e.g., N1, N2, N4, and N9) using\\ \quad proprietary interfaces like EPC\\ - Support network slicing (Slice-based session establishment)\end{tabular} \\ \hline
	\end{tabular}
\end{table*}
\subsection{{Possible States of CN Architecture Towards 5G and Beyond}}
A CSP can have three kinds of states, namely initial, intermediate, and final states when it \textcolor{black}{aims} to migrate from one technology to another. The migration path towards 5GC is very dependent on the initial state of each CSP (Table \ref{core_migration_states}).

One important tip is that the CSP can still maintain the old physical EPCs in int4 and int5 states if the cost of phasing them out is too high. In this case, the EPC+, which only has a presence in a few geographical areas, can serve 5G option 3 to end-users. Moreover, fin1 and fin2 states are adequate for the CSPs that may not want to use EPC+ for migration (int3 and int4 states).

\begin{table*}[]
	\caption{Core Network's different migration states \cite{GSMA,netmanias2019comparison}.}
	\label{core_migration_states}
	\centering
	\scriptsize
	\begin{tabular}{|c|c|l|c|}
		\hline
		\rowcolor[HTML]{34CDF9} 
		\textbf{State} & \textbf{State\#} & \multicolumn{1}{c|}{\cellcolor[HTML]{34CDF9}\textbf{Description}} & \textbf{3GPP Option} \\ \hline
		\cellcolor[HTML]{FFCE93} & ini1 & Physical EPC & Option 1   \\ \cline{2-4} 
		\multirow{-1.75}{*}{\cellcolor[HTML]{FFCE93}\begin{tabular}[c]{@{}c@{}}Initial\\ (2017-2020)\end{tabular}} & ini2 & Virtualized EPC & Option 1  \\ \cline{2-4} 
		\cellcolor[HTML]{FFCE93} & ini3 & Physical and Virtualized EPC (Both serving 4G) &  Option 1 \\ \cline{2-4} 
		\cellcolor[HTML]{FFFE65} & int1 & EPC+ (Serving both 4G and 5G Option 3) & Option 3 \\ \cline{2-4} 
		\cellcolor[HTML]{FFFE65} & int2 & Virtualized EPC (Serving 4G) interworking with EPC+ (Serving 5G Option 3) & Option 3 \\ \cline{2-4}
		\cellcolor[HTML]{FFFE65} & int3 & Virtualized EPC (Serving both 4G and 5G Option 3 (with lower capacity)) & Option 3  \\ \cline{2-4} 
		\cellcolor[HTML]{FFFE65} & int4 & Physical EPC (Serving 4G) interworking with virtualized EPC (serving 5G option 3) & Option 3 \\ \cline{2-4} 
		\multirow{-4}{*}{\cellcolor[HTML]{FFFE65}\begin{tabular}[c]{@{}c@{}}Intermediate\\ (2019-2023)\end{tabular}} & int5 & Physical EPC (Serving 4G) interworking with EPC+ (serving 5G option 3) & Option 3 \\ \cline{2-4} 
		\cellcolor[HTML]{32CB00} & fin1 & Physical EPC (Serving 4G) interworking with 5GC (Serving 5G Options 2/4) & Options 2/3/4\\ \cline{2-4} 
		\cellcolor[HTML]{32CB00} & fin2 & Virtualized EPC (Serving 4G) interworking with 5GC (Serving 5G Options 2/4) & Options 2/3/4 \\ \cline{2-4} 
		\cellcolor[HTML]{32CB00} & fin3 & EPC+ (Serving 4G) interworking with 5GC (Serving 5G Options 2/4) & Options 2/3/4 \\ \cline{2-4} 
		\multirow{-4}{*}{\cellcolor[HTML]{32CB00}\begin{tabular}[c]{@{}c@{}}Final\\ (2020-2025)\end{tabular}} & fin4 & Unified 5GC (Serving both 4G and 5G Options 2/4) & Options 2/4\\ \hline
	\end{tabular}
\end{table*}
\begin{re}
	Some CSPs can reach a final state directly without using any intermediate state {because of the high maturity level of their network architecture}.
\end{re}
\begin{re}
	Wherever we mention EPC (physical or virtualized), CUPS is not featured in the solution. Also, CUPS is the main characteristic of EPC+ solution alongside being virtualized.
\end{re}

\subsection{Migration Paths towards CN with a 5GC}
Herein, we provide two migration paths from Table \ref{core_migration_states} as examples:\\
\textbf{1) Migration path of the Korean CSP KT towards 5G:} Starting with a mature virtualized EPC (i.e., ini2), they recently updated all their core sites to EPC+ (featuring 3GPP CUPS) in order to be able to serve 5G NSA devices (i.e., int1). It is worth {noting} that the user plane functions have been implemented in their edge sites to lower the latency and benefit more from the CUPS feature. The CSP will continue its smooth journey towards 5G SA by implementing 5GC in their core sites and provide interworking with EPC+ (i.e., fin3). KT claims that the CSP will support 4G, 5G NSA, and 5G SA devices in 2020 using this strategy \cite{netmanias2019comparison}.\\
\textbf{2) Possible migration path of a CSP in developing countries towards 5G:} These CSPs may or may not have virtualized their EPCs at the moment. In the first step, they need to implement some virtualized EPC sites if they have not done this step (i.e., ini2 and ini3). The CSP can support 5G NSA devices with a limited coverage in this step. If the CSP has improved its network intelligence (through using SDN and NFV) and has a {mature} orchestration solution, it can go directly for the state fin2 by upgrading its physical EPC sites to the virtualized ones and provide interworking with the new 5GC sites \cite{GSMA} (i.e., fin2).
{Another possible} scenario for the CSPs is to upgrade some of their vEPC sites to the EPC+ ones to improve the scalability and flexibility of their network (i.e., int1, int2, and int5). Afterwards, the CSP can decide between fin3 or fin4 {as its} final state.


%
%

\section{Transport Network Migration}\label{TN Migration}
In this section, the decision options for migration of TNs are provided. Specifically, the options for increasing bandwidth, adaptation to the C-RAN architecture, and implementation of transport SDN (T-SDN) are described.
\subsection{Bandwidth Considerations}
Primary implementations of 5G networks follow the NSA category in which the eMBB services are of most importance. Therefore, it is predictable that the 5G BSs (gNBs) require much more bandwidth to transport the data of 5G users. For example, each of the current 4G BSs injects about 150 Mbps-1 Gbps traffic into TNs in urban areas, while it is predicted that this value be about 600 Mbps-20 Gbps in 2025 \cite{ericsson2018microwave}. Therefore, the existing solutions for data transmission in TNs require a radical change to come up with the increase in the data rates in the coming years.\\
\indent Operators have to decide about the appropriate transport technologies which can provide the bandwidth requirements. This decision is usually impacted by the availability of fiber infrastructure, the deployment of BSs, and the available spectrum for wireless transmission. Since the fiber-based solutions can provide very high data rates with low latency and high availability, the best solutions are fiber-based ones from a performance point of view.
However, utilizing fiber-based solutions is not possible in many cases. For example, in urban areas, the operators may not be allowed to trench the streets for fiber installations. Moreover, the deployment of fiber infrastructure comes with a large investment and results in an increased time to market.

On the other hand, wireless solutions can be easily installed, incur lower costs, and can be utilized in much less time than that of fiber-based ones. Nevertheless, wireless solutions impose some limitations on the transport links in terms of data rate, latency, and availability. The predictions show that the utilization of fiber-based solutions will increase in the coming years, however, the wireless solutions will play an important role in 5G TNs as well \cite{ericsson2018microwave}.

Operators need to enhance fiber-based and wireless solutions for the implementation of 5G networks. In doing so, three options may be identified:
\subsubsection{{Replacement of Wireless Solutions With Fiber-based Ones}}
Regardless of 5G network rollouts, the traffic of 4G networks is increasing and utilizing fiber-based solutions can fulfill the increasing data rate requirement of 4G backhaul networks. Thus, the deployment of fiber infrastructure simultaneously provides better 4G users experience and prepares the TN for 5G rollouts. The primary candidates for the deployment of fiber infrastructure are the 4G BSs (eNBs) in dense urban areas which may already have transport bottlenecks. On the other hand, these BSs are the best locations for implementations of the first gNBs. Therefore, there is no need for extra investment in TNs for the deployment of gNBs. However, as described above, the deployment of fiber infrastructure may require a relatively long time. Thus, the operators need to perform appropriate proceedings in a sufficient time duration before commercial deployments. The studies show that a large number of operators need to evolve their TN at least one to two years before the commercial deployment of 5G networks \cite{heavy2019transport}.
\subsubsection{{Enhancement of Wireless Solutions}}
There are some options to determine which frequency band and wireless technology should be adopted. The selection of a wireless solution depends on various considerations such as the required capacity, latency, and the availability of wireless links as well as the available frequency bands, range of wireless links, and device installation considerations. 
\\\indent
Some studies show that achieving a 10 Gbps data rate for wireless links is of interest to many operators \cite{heavy2019transport}. However, such a high data rate can not be provided by utilizing traditional bands (i.e., sub 40 GHz bands ). Therefore, ETSI standardized some new bands in millimeter-wave including V-band and E-band which can provide up to 10 GHz bandwidth for wireless transport links.
Moreover, studies are undergoing for identification of some bands in higher frequencies, i.e., W-band and D-band.
\\\indent
Another challenge of the utilization of wireless links is the mapping of the required quality of service (QoS) and available frequency bands. 5G networks are supposed to serve services with diverse QoS requirements. On the other hand, wireless links have diverse propagation characteristics that may impact the QoS. Therefore, there is a need to appropriately map each type of transmitted service to a proper frequency band. To do so, ETSI introduced the band and carrier aggregation (BCA) technology which can collect the environment information and map each service to a proper band. Moreover, BCA can extend the range of wireless links by sacrificing availability \cite{etsi2017frequency}.
\subsubsection{{Enhancement of Fiber-based Solutions}}
Gigabit passive optical networks (GPONs) are deployed by many operators so far, however, this generation of PONs can not fulfill both capacity and latency requirements of 5G networks. The NG-PON2 networks are promising candidates for replacement of GPONs, however, the viability of this class of PONs is under question for coming years and new applications (e.g., for fronthaul networks). Therefore, the research and standardization process is undergoing for 50 G/100 G-PONs. In addition to PONs, Ethernet interfaces require to be upgraded for bandwidth increase. Although GbE interfaces are sufficient for most of the existing eNBs, they can not transmit the data volume injected by gNBs. Therefore, these interfaces need to be upgraded to 10 GE/ 25 GE (specifically for fronthaul interfaces).
To summarize, the stepwise proceedings required for increasing the bandwidth of TN links is stated in \eqref{TN_Band_Path}.
\begin{figure*}
	\begin{align}\label{TN_Band_Path}
		\# (\bold{1}): \text{Fiber infrastructure preparation
		}&\bold{\rightarrow} \# (\bold{2}): \text{Upgrading fiber-based \& wireless solutions
		}
		\\
		&\bold{\rightarrow} \# (\bold{3}): \text{Expansion of fiber infrastructure \& Bandwidth procurement at E-band
		}\nonumber
	\end{align}
\end{figure*}
\begin{figure*}
	\begin{align}\label{TN_SDN_P}
		&\# (\bold{1}): \text{Hybrid deployment at national-tier
		}\bold{\rightarrow} \# (\bold{2}): \text{Hybrid deployment at regional-tier
		}
		\\
		&\bold{\rightarrow} \# (\bold{3}): \text{Hybrid deployment at local-tier
		}\nonumber\bold{\rightarrow} \# (\bold{4}): \text{Replacing legacy devices with SDN-enabled ones
		}\nonumber
	\end{align}
\end{figure*}

\subsection{{C-RAN Architecture Considerations}}
The deployment of fronthaul network is the most challenging task of C-RAN deployment.
The conventional interface used by 4G networks is the common public radio interface (CPRI). As CPRI data rate requirement increases linearly with over-the-air bandwidth and number of antennas, it is not scalable for 5G networks which benefit from large bandwidth carriers as well as massive MIMO technology.
Therefore, enhanced CPRI (eCPRI) is introduced for the fronthaul interface of 5G networks which exploits the functional splitting concept to reduce the requirements of latency and data rate compared to CPRI. Moreover, eCPRI utilizes packet-based technologies for transmission such as Ethernet and IP which leads to cost reduction. In addition to the fronthaul interface, the F1 interface, also called midhaul, may be utilized for the connectivity of the DUs and CUs. Furthermore, the NG interface connects the CUs with the 5G core network.
The requirements of data rate and latency of these interfaces are impacted by the selected 3GPP functional splitting options. Option 2 (between radio link control and packet data convergence protocol layers) is standardized for DU-CU functional splitting \cite{itut2018tn5g} (also called higher layer splitting) which results in a latency requirement in the order of milliseconds
and a data rate requirement almost equal to the user data rate. At the same time, there are various options for DU-RU functional splitting (also called lower layer splitting) which result in the latency requirements from tens to hundreds of microseconds. As more functions resided in DU rather than RU, the centralization benefits increase, however, the centralization of functions causes challenges in the fulfillment of fronthaul network requirements. Therefore, there is a trade-off between benefits of centralization and fronthaul network requirements, and hence, operators may need to select an optimal functional splitting option based on their RAN requirements and the quality of their fronthaul network.

The operators need some time to prepare the infrastructure for the deployment of fronthaul networks. Therefore, it is necessary to first determine the scenarios for the implementation of C-RAN in order for evolution from distributed networks. As the benefits of centralization are considerable in dense areas, most operators are willing to start the centralization in these areas. 
The most desirable scenarios for C-RAN deployment are large public venues, outdoor urban areas, and high urban areas.

For a smooth evolution of RAN architecture, it may be desirable for operators to start centralizing their existing 4G networks. Having a 4G C-RAN can facilitate the implementation of a centralized 5G NR. However, the CPRI interface currently utilized for 4G networks is TDM-based which is not consistent with packet-based 5G fronthaul interfaces. To resolve this issue, the IEEE 1914 working group has developed the radio over Ethernet (RoE) standard which determines the encapsulation of digitized radio over Ethernet frames. The RoE includes a CPRI mapper which maps/de-maps CPRI frames into/from Ethernet frames. By utilizing RoE CPRI mapper, the operators can have an Ethernet-based 4G C-RAN which can be readily extended for 5G C-RAN. To sum up, the process of centralization can be started from hotspot scenarios and for 4G networks, then 5G NR BSs can be added to the network when the capacity and latency requirements of fronthaul interfaces can be fulfilled.
\subsection{ {Software-Defined Netwrking Considerations}}
SDN brings many advantages such as flexibility, scalability, and agility for TNs. Since the implementation of the SDN in TNs includes specific considerations, transport SDN (T-SDN) is introduced to enable SDN for TNs. Also, the T-SDN is one of the enablers of network slicing. Network slices are realized by coordination of the TN management system and E2E management system.
As recommended by \cite{itut2018tn5g}, the TN management system can be implemented by an SDN controller.

To evolve legacy TNs for supporting the T-SDN, both devices and architecture of TNs should be upgraded. In the following, the migration process of TN devices (i.e., forwarding/routing devices) and TN architecture is described.
\subsubsection{{ Migration of Transport Network Devices}}
The migration working group of open networking foundation (ONF) suggests three types of network devices that facilitate the TN migration: 1) Legacy devices: the conventional network devices with integrated CP and DP; 2) SDN-enabled devices: the SDN-enabled switches with decoupled CP from DP, and CP residing external to the device; and 3) Hybrid devices: devices with both legacy DP and CP and SDN capabilities (most of the legacy devices are upgradable to be SDN-enabled via software/hardware upgrades).\\
\indent There are three deployment approaches for migrating legacy networks to SDN-enabled networks: 1) Greenfield deployment: direct upgrade of the network devices to the SDN-enabled ones. The greenfield deployment is appropriate when there is no legacy network or there is a legacy network that is directly upgraded; 2) Mixed deployment: co-existence of SDN-enabled with legacy devices. In this approach, the SDN Controller and the legacy devices will need to exchange routing information between each other via the legacy CP agents; and 3) Hybrid deployment: co-existence of legacy, hybrid, and SDN-enabled devices. In the hybrid deployment, the upgradable legacy devices can be upgraded to hybrid devices and gradually the remaining of the legacy devices can be decommissioned when the new SDN-enabled devices are deployed.
\subsubsection{{ Migration of Transport Network Architecture}}
The TNs are likely to be multi-administrative, multi-layer and widely dispersed. Therefore, TNs are usually divided into multiple domains, for example, domains of different infrastructure providers, domains of optical and packet layers, and domains of different geographical regions. To evolve such TNs to T-SDN enabled networks, the requirements of different domains should be met. To resolve this issue, a hierarchical architecture can be adopted
in which some domain SDN-controllers (also called as child controllers) manage the devices in their domain, and a parent SDN-controller manages the E2E network (which can be inferred as the previously mentioned transport management system). By a domain-based evolution of TNs, the migration process can be performed smoothly. For example, the first step of the migration might be the evolution of a national domain for the 5connection of  multiple CN functions and/or multiple data centers around a country. Then, the lower layer domains (i.e., regional and local domains) can gradually evolve to be SDN-enabled. Moreover, for the intra-domain evolution,  each of the deployment approaches (i.e., greenfield, mixed, and hybrid) can be adopted based on the stats-quo of the operators' TN.
The required steps for deployment of T-SDN are outlined in \eqref{TN_SDN_P}.

\section{conclusion remarks}\label{conclusion}
\textcolor{black}{
This paper provided a practical and comprehensive view on how to migrate to 5G and beyond. We introduced and discussed different migration paths in RAN, TN, and CN considering various factors such as marketing and technical factors.  More importantly, we explained and considered the E2E view of the network that comprises all the domains of a telecom network and gave E2E solutions towards 5G and E5G. This paper is helpful for both academia and industry (e.g., operators) to have information about  5G with an eye on what would happen beyond the standardization and deployment perspectives.  
}

\bibliography{citation_Migiration_Final}
\vspace{1em}	
\bibliographystyle{ieeetran}
  	{\textbf{Abulfazl Zakeri}}
	 is currently pursuing the Ph.D. degree in the Department of Electrical and Computer Engineering, Tarbiat Modares University, Tehran, Iran. 
	 His current research interests include wireless communication networks with emphasis on softwarization, E2E
	 network slicing, optimization, and neural networks.
	 \\\\\\
	{\textbf{Narges Gholipoor}}
	is currently pursuing the Ph.D. degree in the Department of Electrical and Computer Engineering, Tarbiat Modares University, Tehran, Iran. 
\\\\\\
	{\textbf{Mohsen Tajalifar}}
	is currently pursuing the Ph.D. degree in the Department of Electrical and Computer Engineering, Tarbiat Modares University, Tehran, Iran. 
	\\\\\\{\textbf{Sina Ebrahimi}}
	is currently a research assistant in the Department of Electrical and Computer Engineering, Tarbiat Modares University, Tehran, Iran and has graduated from his M.Sc. in software engineering from the same university. 
\\\\\\
	{\textbf{Mohammad Reza Javan}}
	is an associate Professor at the Department of Electrical and Computer at Shahrood University, 
Shahrood, Iran. His research interests include design and analysis of wireless communication networks, with an emphasis on the application of optimization theory.
	\\\\\\
	\textbf{Nader Mokari}
	is an associate professor at the Department of Electrical and Computer at Tarbiat Modares University, 
	Tehran, Iran.  His research interests include design,
	analysis, and optimization of communication
	networks.
\\\\\\	{\textbf{Ahmadreza Sharafat}}
	is a professor at the Department of Electrical and Computer at Tarbiat Modares University, Tehran, Iran.
\end{document}